 \documentclass[final,1p,times,twocolumn]{elsarticle}

\usepackage{graphics}
\usepackage{graphicx}
\usepackage{epsfig}
\usepackage[latin1]{inputenc} 
\usepackage{amssymb}
\usepackage{amsthm}

\begin{document}

\begin{frontmatter}



\title{Two-dimensional lattice polymers: adaptive windows simulations}


\author[ufmg,ufg]{A. G. Cunha-Netto}
\ead{agcnetto@fisica.ufmg.br}
\author[ufmg]{Ronald Dickman}
\ead{dickman@fisica.ufmg.br}
\author[ufg]{A. A. Caparica}
\ead{caparica@fis.ufg.br}


\address[ufmg]{Departamento de F\'{\i}sica, Instituto de Ci\^encias Exatas, Universidade Federal de Minas Gerais, C.P.702, 30123-970 Belo Horizonte, Minas Gerais, Brazil}
\address[ufg]{Instituto de F\'{\i}sica, Universidade Federal de Goi\'as, C.P. 131, 74001-970, Goi\^ania, Goi\'as, Brazil}

\begin{abstract}
We report a numerical study of self-avoiding polymers on the square lattice, including
an attractive potential between nonconsecutive monomers. Using Wang-Landau
sampling (WLS) with
adaptive windows, we obtain the density of states for chains of up
to $N=300$ monomers and associated thermodynamic quantities.
The method enables one to simulate accurately the low-temperature regime, which is virtually
inaccessible using traditional methods. Instead of defining fixed
energy windows, as in usual WLS, this method uses windows with boundaries that
depend on the set of energy values on which the histogram is flat at a given stage
of the simulation. Shifting the windows each time the modification factor $f$ is
reduced, we eliminate border effects that arise in simulations using fixed windows.

\end{abstract}


\end{frontmatter}

\section{Introduction}

In recent years the Wang-Landau sampling (WLS) \cite{landau1,landau2} has been applied to many systems and has become a
well established Monte Carlo algorithm. Like
the Metropolis algorithm, it is applicable to almost all stochastic simulations.
In particular the method has been used in studies of polymers \cite{3d,jain,rampf_binder_paul,parsons_williams}
and proteins \cite{rathore,wuest_landau}.

One of the important features of the WLS is that in general one can
simulate larger systems than with conventional Monte Carlo algorithms. To
do this one splits the total energy range into slightly overlapping subintervals, so called windows, and
simulates each separately. The density of states $g(E)$ for the whole energy space is then obtained by multiplying
the density of states in each window by an appropriate factor which assures continuity of the function.
Such a result is sufficient to calculate canonical averages. Nevertheless some distortions arise at the borders
of the energy windows. Limiting the energy space causes some distortions in the estimation of the density of states.
Recently a way of circumventing this problem, using adaptive windows \cite{adaptive_window} in WLS was developed.
The method consists in dividing the parameter space into intervals during the simulation. The segments are
created using a mobile edge where the border position depends on the portion of the histogram that has already
become flat. As a result the density of states, the probability distribution and the thermodynamic properties
do not suffer from the distortions that arise using conventional WLS with fixed windows.

The remainder of the paper is organized as follows. In Section 2, we briefly present the model and the
evolution protocol. In Section 3, we discuss some anomalies in the probability distribution and critical
temperature (obtained from the maximum of the specific heat) that arise using fixed windows.
In Section 4 we describe the adaptive windows (AW) algorithm, and in Section 5 present results of AW simulations
including the \textit{entire} range of energies for chains up to 300 monomers on the square lattice.

\section{The model}

We simulate a lattice polymer consisting of $N$ monomers; the polymer may assume any self-avoiding walk (SAW)
configuration on a two-dimensional lattice. In addition to the SAW condition, which represents excluded volume,
we include a monomer-monomer
attraction \cite{baumgartner}: each pair of nonbonded nearest-neighbor monomers contributes the amount
$-\varepsilon$ to the energy. Due to this attractive potential, the typical configuration
changes from an open ``coil'' to a dense ``globule'' at the collapse temperature $T_c$.
The Hamiltonian of the system can be written as

\begin{equation}
\mathcal{H}=-\varepsilon\sum_{<i,j>}\sigma_{i}\sigma_{j},
\label{hamiltoniana}
\end{equation}
where $\sigma_{i}=1$ $(0)$ if the site $i$ is occupied (vacant), and the sum is
over nearest-neighbor pairs.



We sample the configuration space using reptation dynamics \cite{rept2,rept3}, which consists in randomly selecting
one of the ends of the chain and transferring a monomer from one end to the other end at random.

\section{Systematic errors}

In order to simulate larger systems using WLS, the authors of the method suggest splitting the total energy range
into slightly overlapping subintervals and simulating each separately. The density of states $g(E)$ for the whole
energy range is then obtained by multiplying the density of states in each window by an appropriate factor, to
impose continuity of the function. Nevertheless some distortions arise at the borders of the energy intervals
suggesting systematic errors. Schulz \textit{et al.}\cite{schulz} introduced a new rule aimed at correcting this problem:
update the current energy value in $g(E)$ and the energy histogram whenever a move is rejected because its energy
is greater than the maximum allowed for the window. The procedure partially corrects the density of states,
but some distortions remain in its derivatives. These difficulties can easily be seen in the energy probability
distribution $P(E)=g(E) exp[-E/k_BT]$. In Fig.(\ref{fig:prob_1seed}) we show results for the distribution of
probabilities at temperatures $T=0.7, 1.25$ and $2.5$ for a chain of $N=200$ units. The simulation was performed using
four windows, spanning the range from $E_{min}=-172$ to $E_{max}=0$.

\begin{figure}[!htb]
\centering
\includegraphics[clip,angle=0,width=0.9\hsize]{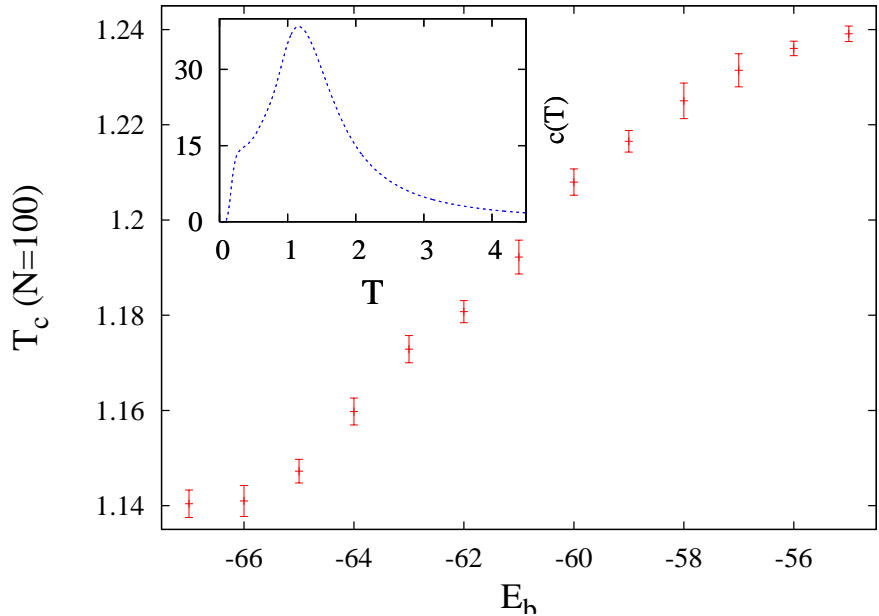}
\caption{\textit{Dependence of the maximum of the specific heat on the position of the border between two windows.
The inset shows a typical result for the specific heat.}}
\label{fig:tceb}
\end{figure}

Errors in the probability distribution naturally induce errors in the specific heat (see Ref. \cite{adaptive_window}).
In order to characterize these errors we carried out WLS simulations using two windows
and estimated the specific heat (obtained as usual from the variance of the energy) as we vary the position of the
border, for a chain of $N=100$ monomers (see Fig.(\ref{fig:tceb})).
(The simulations were performed using a flatness criterion of $80$\%, that is, $H(E)>0.8\overline{H}$
for all energies in the window of interest, where the overline denotes an average over energies.
Simulations were halted when $f \sim 1 + 10^{-7}$. Uncertainties were estimated using ten independent runs.)

We tried to circumvent this problem using multiple ``random walkers'' to construct the density of states.
We simulated ten polymers in two different situations: (i) all the chains running through the whole energy space and
(ii) the chains restricted to different energy intervals with large overlaps. In case (i) good results where obtained
for small chains, but for larger polymers the simulation does not converge. The motivation behind procedure (ii) is that
the deficiencies associated with a walker close to the border might be corrected by another walker
in a different interval, but this evidently did not occur; the resulting density of states includes
small discontinuities at the borders.

We therefore conclude that WLS with fixed windows leads to distorted results for the density of states $g(E)$,
when applied to lattice polymers. Similar distortions were observed in studies of the Potts \cite{adaptive_window}
and Blume-Capel \cite{blumecapel,blumecapel2} models using WLS with fixed windows for the estimation of the joint
density of states.

\section{The method}

The idea of the adaptive windows algorithm derives from the observation that during the WLS of a given system the
histogram becomes flat gradually, proceeding from higher energy levels to lower ones. It is therefore possible
in the course of the simulation to identify an interval of the energy space where the histogram is already flat
according to the usual criterion, and so to set the window on this interval.
Simulation of this interval can then be suspended, while the rest of the spectrum is sampled, allowing a small overlap
with the already defined window.  Subsequently new windows can be defined and closed following the same
procedure until one containing the ground state energy is created. Before moving to the next stage,
with a new modification factor $f$, we impose continuity on
the current density of states $g(E)$ following the procedure described above.
The simulation with the new factor $f$ again runs through the entire spectrum, providing the correction of
possible defects arising from the borders. The procedure described above again generates a sequence of windows. Here
an important restriction must be imposed: do not permit a border to be defined at (or very near) the position
of a boundary formed at the previous stage. Without this rule, the method would suffer from the same deficiencies as
fixed window WLS.

In brief, we begin a WLS as usual setting the initial values of the density of states (histogram) as 1 (0) for all levels
of energy. The random walk in the energy space runs through all energy levels from $E_{min}$ to $E_{max}$ with a
probability $p(E\rightarrow E')=g(E)/g(E')$, where $E$ and $E'$ are the energies of the current and the new possible
configurations. Whenever a configuration is accepted we update $H(E')\to H(E')+1$ and $g(E')\to g(E') \times f_i$,
where $f_0=e$ and $f_{i+1}=\sqrt{f_{i}}$. After $N_{1}$ Monte Carlo steps (in practice we use $N_{1}=10^{4}$) we check
the histogram for the flatness criterion on a minimal window, of width $W=(E_{max}-E_{min})/10$, in the upper portion of
the spectrum. If it is not flat, we perform an additional $N_{1}$ Monte Carlo steps, and check again,
repeating until the histogram
is flat on the minimal window. Once this condition is satisfied, we check whether the histogram is in fact flat on a larger
interval. This is done by adding, one by one, the levels below the minimal window, calculating $\overline{H}$, and checking
for flatness. As a result we identify the largest window over which flatness is satisfied from $E^{*}$, the last level for
which the histogram was still flat, to $E_{max}$. Let
$\Delta E$ be the overlap between two adjacent windows (usually $\Delta E=3$ for lattice polymers). Then the matching level between the first and the second window will be $E_{1}=E^{*}+\Delta E$ and the random walk in the next stage of the
simulation will be restricted to energies $E_{min}\leq E\leq E_{1}+\Delta E$. Again we run the simulation until the
flatness criterion is satisfied in a minimal window and identify the largest window over which flatness is still satisfied
from a new $E^{*}$ to $E_{1}+\Delta E$, define the new matching level $E_{2}=E^{*}+\Delta E$ between the second
and the third window and proceed as above, until all possible energies have been included in a window with a flat
histogram, with the precaution that the final window with lower limit $E_{min}$ always has a width $\geq W$.

\begin{figure}[!hbt]
\centering
\includegraphics[clip,angle=0,width=0.8\hsize]{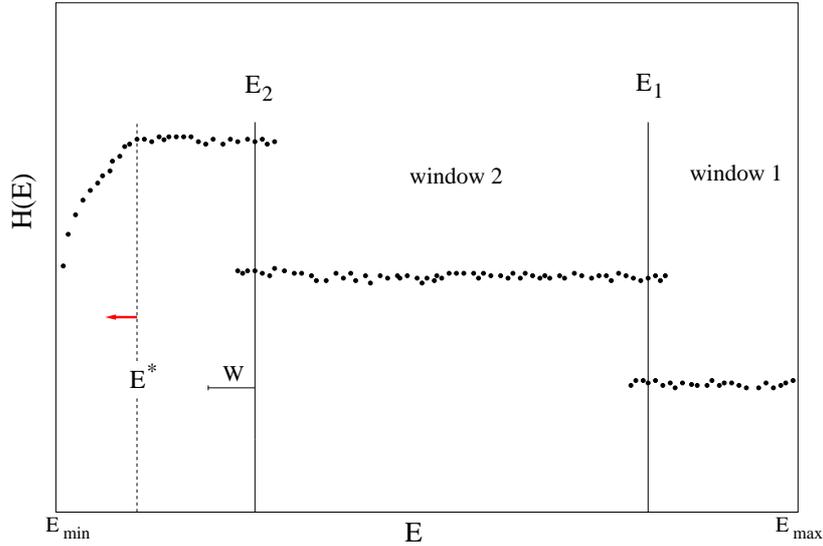}
\caption{Schematic of adaptive windows;
the value of $E_{max}$ depends on the model, while $E^*$ and $E_1$,
$E_2$, etc., are determined during the simulation not fixed beforehand. }
\label{fig:scheme}
\end{figure}

Fig.(\ref{fig:scheme}) is an illustration of how the windows are formed. $E_1$
and $E_2$ are the matching levels of two already established windows and $E^*$ is the
limit of the third window being closed.

Once all windows have been formed with the last one beginning from the ground state
$E_{min}$, we impose continuity to the current density of states by equalizing $g(E)$
of contiguous windows at the matching levels $E_1,E_2,E_3,..,E_n$. The modification
factor is then updated $f \to \sqrt f$ and the random walk is restarted with
the system free to visit all energy levels from $E_{min}$ to $E_{max}$. It is
important to stress at this point that if a new window ends very close to the limit
of a window of the former modification factor, we avoid this vicinity by reducing its
size by, say, $0.25W$ levels. This way we avoid the systematic errors observed in
WLS with fixed windows. The simulation continues
until $f$ is very close to unity, e.g., $f\approx 1+10^{-7}$.

The method can be easily generalized to systems with a multiparametric density of states
$g({\cal X}_1,{\cal X}_2,...,{\cal X}_n)$. Simulations of the Potts and Blume-Capel
models \cite{adaptive_window,blumecapel2} yielded excellent results with random walks
in a two-dimensional parameter space. In both cases one parameter is maintained
unrestricted and the windows are formed in the other variable.

\section{Results}

\begin{figure}[!ht]
\centering
\includegraphics[clip,angle=0,width=1.0\hsize]{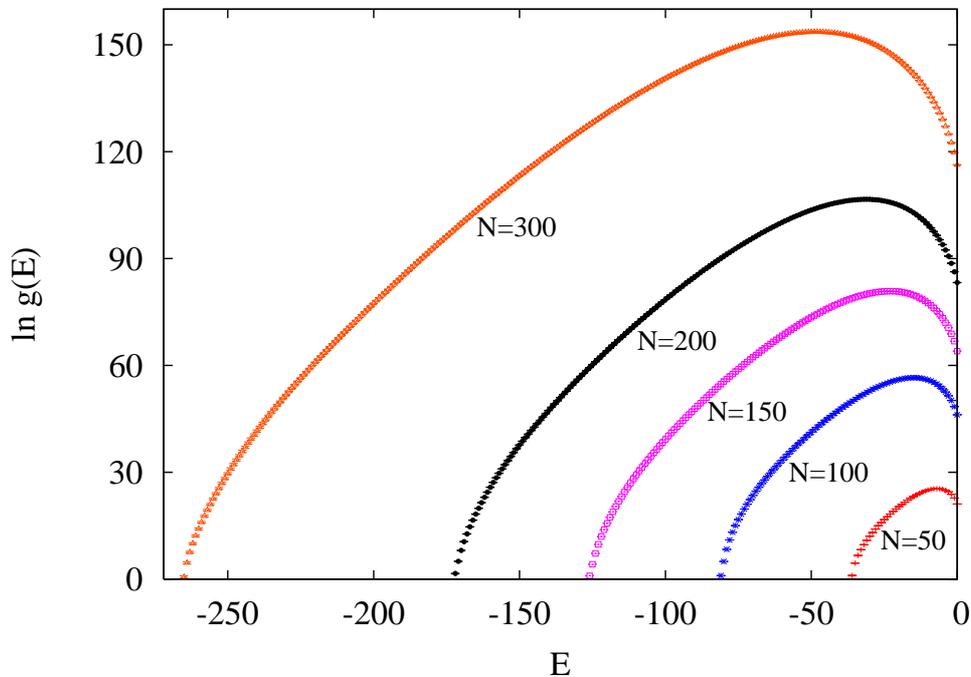}
\caption{Density of states of two-dimensional polymers using adaptive windows scheme.}
\label{fig:lng}
\end{figure}

Improving WLS with adaptive windows, we can simulate larger system sizes without border effects.
In Fig.(\ref{fig:lng}) we show the density of states of chains of up to N=300 monomers including the entire range of energies.
(As is known, some low-energy configurations are inaccessible to reptation; we believe that the error incurred
is far bellow the precision of our simulation.)

\begin{figure}[!ht]
\centering
\includegraphics[clip,angle=0,width=0.9\hsize]{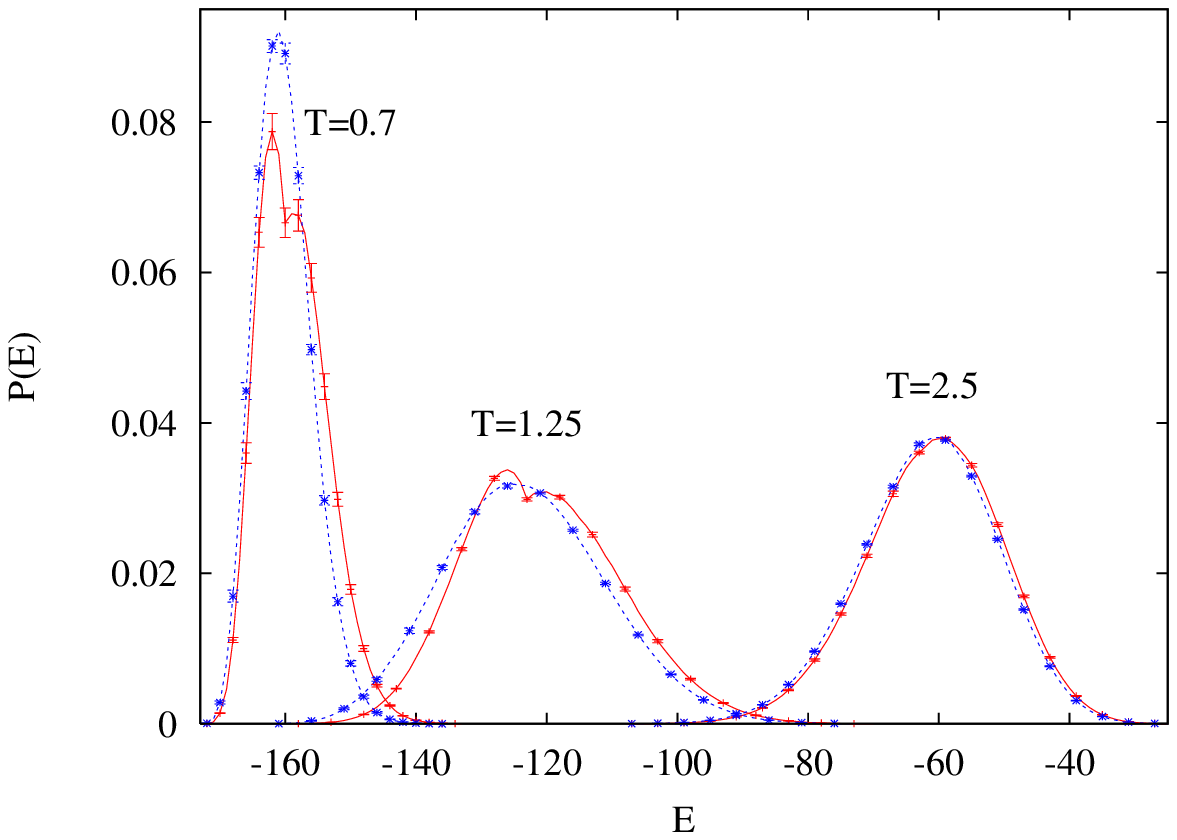}
\caption{\textit{Probability distribuition normalized by the partition function for a chain of $N=200$ monomers. Solid line: results using WLS with fixed windows. The
spectrum was split into four windows with borders at $E_b=-160,-123,-63$. Dashed line: results using adaptive windows. Only a few typical error bars are shown for ten independent runs.}}
\label{fig:prob_1seed}
\end{figure}

It is important to point out that WLS of polymers of such sizes does
not converge without the use of windows. In Fig.(\ref{fig:prob_1seed}) we show results
for the probability distribution obtained from WLS with fixed and with adaptive
windows. Discontinuities like these affect the thermodynamic quantities such as the
specific heat, yielding unreliable results.

\begin{figure}[!hb]
\centering
\includegraphics[clip,angle=0,width=0.9\hsize]{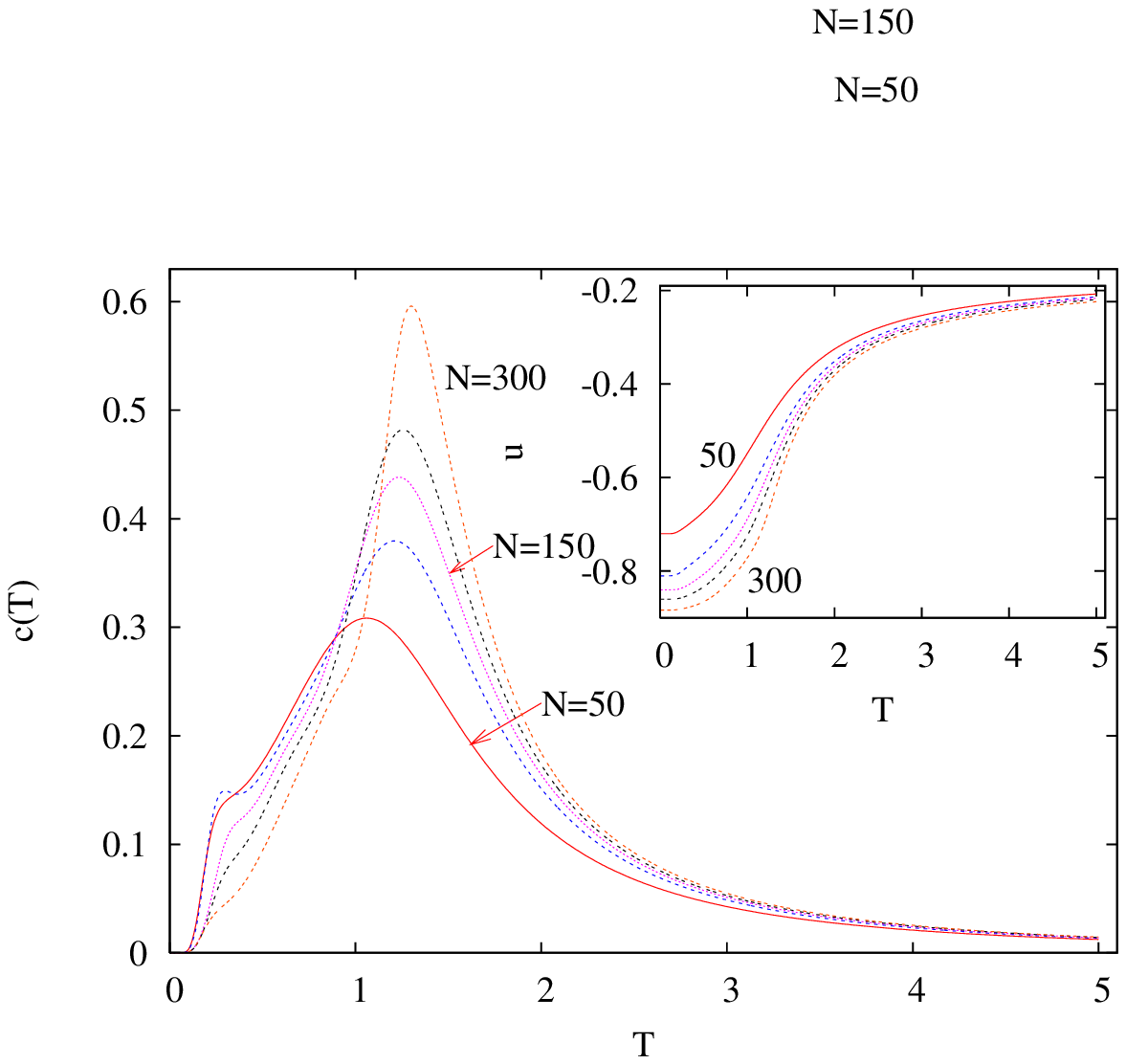}
\caption{(color online) Specific heat per monomer versus temperature for $N=50, 100, 150, 200$ and $300$. In the inset we show the internal energy versus temperature.}
\label{fig:cv}
\end{figure}

In Fig. (\ref{fig:cv}) we show the results for the specific heat as a function of temperature. The inset shows the internal energy versus temperature. We believe that the maxima in specific heat at low temperatures represent a surface effect that will disappear as N $\to \infty$, as observed in studies of lattice animals \cite{animals}.

Average values are calculated using ten independent runs; error bars are smaller than the symbols.

\section{Conclusions}

We study polymers on a square lattice using Wang-Landau sampling with adaptive windows.
In this case, splitting the energy spectrum and simulating each window separately, as in conventional WLS,
does not yield reliable results.
Using the adaptive windows scheme we eliminate this problem by forcing the window positions to shift
during the simulation. We determine the density of states, probability distribution, internal energy and
specific heat for chains of up to $N=300$ monomers, over the entire range of energies.

\section*{acknowledgments}
The authors would like to acknowledge S.-Ho Tsai and D. P. Landau for many interesting discussions and several helpful suggestions. We are grateful to Brazilian science agencies CAPES, CNPq, FUNAPE-UFG and Fapemig for financial support.

\end{document}